\documentclass [prb,twocolumn,superscriptaddress,preprintnumbers] {revtex4}
\usepackage[]{amssymb,amsmath,amsfonts}
\usepackage[]{graphicx}

\begin{document}
\title{Temperature dependence of the thermal boundary resistivity of glass-embedded metal nanoparticles}

\author{Francesco Banfi}
\email{francesco.banfi@dmf.unicatt.it}
\affiliation{i-LAMP and Dipartimento di Matematica e Fisica, Universit$\grave{a}$ Cattolica, I-25121 Brescia, Italy}
\affiliation{FemtoNanoOptics Group, LASIM, Universit\'e Lyon 1, CNRS, 69622 Villeurbanne, France}

\author{Vincent Juv\'e}
\altaffiliation[Present address: ]{Max-Born-Institut f\"{u}r Nichtlineare Optik und Kurzzeitspektroskopie, 12489 Berlin, Germany}
\affiliation{FemtoNanoOptics Group, LASIM, Universit\'e Lyon 1, CNRS, 69622 Villeurbanne, France}

\author{Damiano Nardi}
\altaffiliation[Present address: ]{JILA, University of Colorado at Boulder, Boulder, Colorado 80309, USA}
\affiliation{i-LAMP and Dipartimento di Matematica e Fisica, Universit$\grave{a}$ Cattolica, I-25121 Brescia, Italy}

\author{Stefano Dal Conte}
\altaffiliation[Present address: ]{Department of Applied Physics, Eindhoven University of Technology, 5600 MB Eindhoven, The Netherlands}
\affiliation{Dipartimento di Fisica A. Volta, Universit\`a di Pavia, I-27100 Pavia, Italy}

\author{Claudio Giannetti}
\affiliation{i-LAMP and Dipartimento di Matematica e Fisica, Universit$\grave{a}$ Cattolica, I-25121 Brescia, Italy}

\author{Gabriele Ferrini}
\affiliation{i-LAMP and Dipartimento di Matematica e Fisica, Universit$\grave{a}$ Cattolica, I-25121 Brescia, Italy}

\author{Natalia Del Fatti}
\affiliation{FemtoNanoOptics Group, LASIM, Universit\'e Lyon 1, CNRS, 69622 Villeurbanne, France}

\author{Fabrice Vall\'ee}
\affiliation{FemtoNanoOptics Group, LASIM, Universit\'e Lyon 1, CNRS, 69622 Villeurbanne, France}

\date{\today}

\begin{abstract}
\noindent 
The temperature dependence of the thermal boundary resistivity is investigated in glass-embedded Ag particles of radius 4.5~nm, in the temperature range from 300 to 70 K, using all-optical time-resolved nanocalorimetry. The present results provide a benchmark for theories aiming at explaining the thermal boundary resistivity at the interface between metal nanoparticles and their environment, a topic of great relevance when tailoring thermal energy delivery from nanoparticles as for applications in nanomedicine and thermal management at the nanoscale.
\end{abstract}

%
\maketitle
\clearpage
\indent With the ever decreasing size of nanodevices, investigation and modeling of heat exchange at the nanoscale has become of central technological interest. Under a fundamental standpoint, metal nanoparticles (NP) embedded in a host matrix constitute a model system, as they can be selectively heated-up and their cooling monitored using time-resolved spectroscopy \cite{Hartland2011, Juve2009, Plech2004}. Furthermore, the thermal dynamics occurring between an optically excited metal nanoparticle and the surrounding environment is of direct relevance for a variety of applications ranging from photothermal cancer therapy\cite{Irsch2003, El-Sayed2007} and selective drug delivery,\cite{Paasonen2007} to thermoacoustic imaging and electromagnetic waveguiding in dielectric-embedded plasmonic devices.\cite{Rini2005} The electromagnetic energy harvested by the NP is dissipated as thermal energy in the environment. The corresponding energy flux $J_{p}$ is ruled by the thermal boundary resistivity $\rho_{bd}$, i.e., Kapitza resistivity, and by the temperature mismatch $\Delta T$ between the two media: $J_{p}$=$\Delta T$/$\rho_{bd}$. Investigating $\rho_{bd}$ is therefore crucial to tailor thermal energy delivery from the NP to the matrix or matrix-embedded target and, more generally, to analyze heat transfer at the nanoscale.\\
\indent Whereas effort has been devoted to understand and model the Kapitza resistivity between two solids, both bulk and thin films\cite{Swartz1989, Stoner1993, Cahill2003}, the scenario remains relatively unexplored when one of the two materials downscales to the nanometer range. Several confinement effects may modify $\rho_{bd}$, for instance, as the dimension of the NP becomes comparable to the thermal diffusion length of the host material\cite{Siemens2010, Chen2001, Palpant2008}, or is reduced to a point where the continuum solid approximation to the elastic problem becomes questionable. A lack of extensive experimental evidence\cite{Juve2009, Plech2004PRB, Wilson2002} spanning the space of parameters affecting $\rho_{bd}$, most notably the temperature\cite{Swartz1989, Stoner1993}, has so far prevented a consistent account of the mechanisms ruling the Kapitza resistivity at the nanoscale. When confronted with the problem of measuring the heat transfer from a nanoscale object, the challenges stand in: (a) a probe speed requirement, dictated by the fact that the time for heat exchange between the sample and the thermal reservoir decreases with the decreasing sample's mass; (b) a non-contact probe requirement, to avoid the addendum heat capacitance contribution from the probe itself.\\
\indent In this Letter we use time-resolved all-optical nanocalorimetry\cite{Banfi2010} to overcome such challenges and investigate the temperature dependence of the cooling dynamics of glass-embedded Ag particles of radius $R=4.5$ nm. The Kapitza resistivity is shown to increase by a factor of two with decreasing temperature from 300 to 70 K, a trend consistent with existing models.\\ 
\begin{figure}[t]
\centering
\includegraphics[keepaspectratio,clip,width=1\columnwidth]{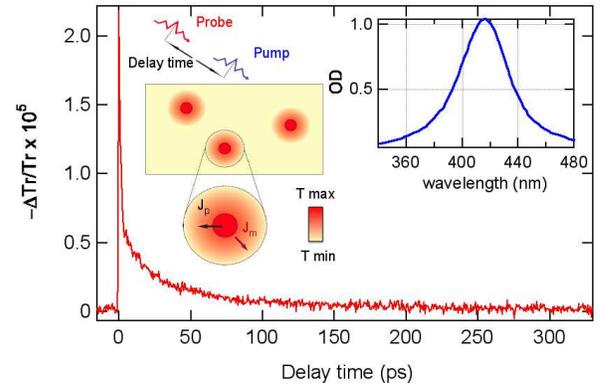}
\caption{(Color online) Measured time-resolved relative transmission change for $T_{cryo}=200$ K. The pump and probe pulses have wavelengths respectively at 400 nm and 800 nm. In the cartoon the thermal fluxes $J_{p}$ and $J_{m}$ are represented together with the temperature profile within the sample. Inset: measured OD of the sample outlining the Ag nanoparticles' LSPR.}
\label{Figure1_3}
\end{figure}
\indent The Ag nanospheres are embedded in a 50\% BaO - 50\% P$_{2}$O$_{5}$ glass matrix of thickness $L=50$~$\mu$m. The metal volume fraction is $2\cdot10^{-4}$. The sample was synthesized using a fusion and heat treatment technique.\cite{Uchida1994, Nelet2004} The samples' optical density (OD) shows enhanced absorption in the blue portion of the spectrum due to the localized surface plasmon resonance (LSPR) of the Ag NP - see inset of Fig.~\ref{Figure1_3}.\\
\indent The time-resolved measurements were performed using a Ti:Sapphire cavity dumped oscillator - 800~nm wavelength, 120~fs pulse temporal width at full width half maximum. The Ag NPs are selectively excited by the frequency-doubled pulse at 400~nm wavelength close to the LSPR in order to maximize energy absorption in the particle. This leads to a fast heating of the electrons of the NPs that thermalize with the lattice on a few picoseconds time-scale, the thermal energy being subsequently delivered to the matrix. Care was taken to minimize average heating \cite{Giannetti2007} of the glass matrix by keeping the energy per pulse as low as possible while granting a detectable transmission variation. To this end the laser repetition rate was tuned to 540~kHz by means of a cavity dumper. 
Accounting for transmission losses along the optical path, the energy density per pump pulse on the sample surface was $I_{0}\sim0.5$~J/m$^{2}$, while the energy density absorbed per particle per pulse was $U_{V}\sim4\cdot10^{7}$~J/m$^{3}$. The cooling dynamics of the hot NPs to the glass matrix is then followed by measuring the relative change in transmission across the sample $\Delta$Tr/Tr of a time-delayed 800 nm probe pulse. Probing out of the LSPR grants proportionality between the experimental signal and the NP¡Çs temperature rise\cite{Juve2009} (at the expense of the signal amplitude).\\
\indent A typical experimental trace is reported in Fig.~\ref{Figure1_3} for a cryostat temperature $T_{cryo}=200$ K. After excitation by the pump pulse and internal electron-lattice thermalization, i.e., after 6 picoseconds (this step has been extensively investigated in these systems \cite{Hartland2011, Arbouet2003} and will not be discussed here), the signal decay reflects cooling of the hot NP to the matrix. This is governed by the thermal flux $J_{p}$ and $J_{m}$ from the NP to the matrix and from the matrix portion adjacent to the NP to the rest of the matrix, respectively. Considering these two processes, the energy balance is governed by:
\begin{eqnarray} 
&&C_{p}\partial_{t}T_{p}(t)=-\frac{3}{R\rho_{bd}}[T_{p}(t)-T_{m}(R,t)]\label{EquationSystem_p}\\
&&C_{m}\partial_{t}T_{m}(r,t)=\Lambda_{m}r^{-1}\partial_{r}^{2}[rT_{m}(r,t)]\label{EquationSystem_m}
\end{eqnarray}
$T_{p}$ being the NP's temperature, assumed as constant throughout the particle volume, $T_{m}$ the matrix' temperature, $C_{p}$ and $C_{m}$ the particle's and matrix's specific heat per unit volume respectively, and $\Lambda_{m}$ the matrix' thermal conductivity. For the case of \textit{constant} thermal parameters, the temperature increase for the NP, $\Delta T_{p}(t)$, and for the matrix portion in contact with it, $\Delta T_{m}(R,t)$, are analytically accessible working in Laplace space \cite{Jaeger:book} and read:
\begin{eqnarray} 
&&\Delta T_{p}(t)=\int_0^{\infty} \! duf(u,t) \label{DeltaTp}\\
&&\Delta T_{m}(R, t)=\int_0^{\infty} \! du\left[1-\frac{u^{2}}{kgR}\right]f(u,t)
\label{DeltaTm}
\end{eqnarray}
where
\begin{equation} 
f(u,t)=\frac{2k(Rg)^{2}\Delta T_{0}}{\pi}\frac{u^{2}\exp(-\kappa u^{2}t/R^{2})}{[u^{2}(1+Rg)-kRg]^{2}+(u^{3}-kRgu)^{2}},\\
\label{f}
\end{equation}
$\Delta T_{0}$ is the initial temperature increase \cite{DeltaT0} and $\kappa=\Lambda_{m}/C_{m}$, $k=3C_{m}/C_{p}$, $g=1/\Lambda_{m}\rho_{bd}$. In our analysis the thermal resistivity $\rho_{bd}$ is set as a fit parameter, together with $\Lambda_{m}$, which is not precisely known for our glass material.\\
\indent As low temperatures are investigated (in particular, around and below the NP's Debye temperature,  $T_{D}\sim215$ K for Ag), $C_{p}$ and $C_{m}$\cite{ThermalData} cannot be set to their $T_{cryo}$ value and regarded as constant over the particle and matrix's temperature excursion taking place during the experiment. The solution of Eq.s \ref{EquationSystem_p} and \ref{EquationSystem_m}, taking into account the temperature dependent specific heats, is then retrieved iteratively. Fitting is performed starting at a time-delay of 6 ps and setting $T_{m,0}=T_{cryo}$ and $T_{p,0}=T_{cryo}+\Delta T_{0}$. The corresponding values $C_{p}(T_{p,0})$ and $C_{m}(T_{m,0})$ are inserted in Eq.s \ref{DeltaTp} and \ref{DeltaTm}, and the new temperatures $T_{p,1}=T_{cryo}+\Delta T_{p,1}$ and $T_{m,1}=T_{cryo}+\Delta T_{m,1}$ are calculated and adopted in the subsequent time step. The procedure is iterated to reach the maximum experimental time-delay of 320 ps. Values for $\rho_{bd}$ and $\Lambda_{m}$ are obtained maximizing the likelihood between the theoretical $T_{p}(t; \rho_{bd}, \Lambda_{bd})/\Delta T_{0}$ and experimental $-\Delta$Tr/Tr traces.\\
\begin{figure}[t]
\includegraphics[keepaspectratio,clip,width=0.95\columnwidth]{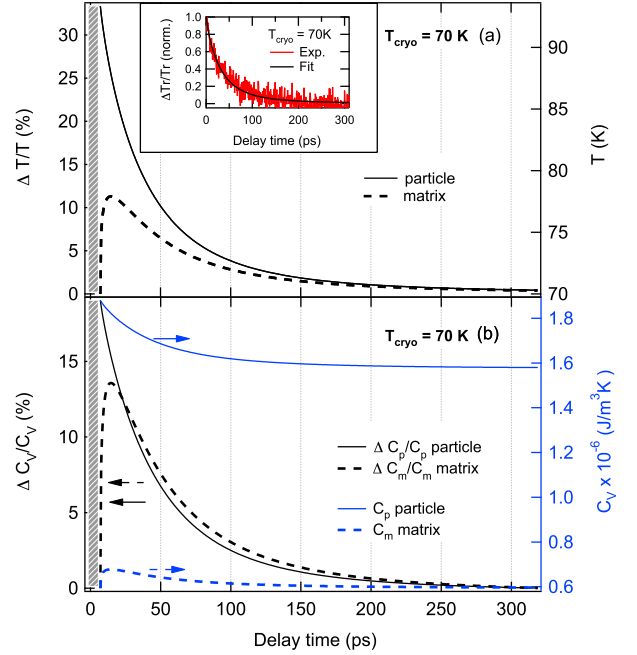}
\caption{(Color online) Time evolution of temperature and specific heat for $T_{cryo}=70$ K. Panel (a): relative temperature variation (left axis) and absolute temperature (right axis) of the NP (full line) and of the adjacent matrix (dashed line). Inset: experimental transmission change normalized to the value at 6 ps (red curve) and its best fit (black curve). Panel (b): relative specific heat variation (left axis) and absolute specific heat (right axis) of the NP (full line) and of the adjacent matrix (dashed line).}
\label{Figure2_3}
\end{figure}
\indent The resulting dynamics of temperatures and specific heats are exemplified in Fig.~\ref{Figure2_3} (a) and (b) respectively for the lowest studied temperature.
The internal thermalization of the NP is achieved at $T_{p}=93$ K. As time evolves the NP cools down, increasing $T_{m}(R)$, the maximum value of $T_{m}(R)$ being attained at $t\sim15$ ps. For longer time-delays both $T_{p}$ and $T_{m}(R)$ decay toward the asymptotic value $T_{cryo}$. The same trend applies to the specific heats $C_{p}$ and $C_{m}(R)$, showing maximum relative changes during the experiment in the 15-20\%  range. A similar behavior is obtained for measurements performed at higher temperatures, although with a smaller excursion amplitude (as indicated by the arrows in Fig.~\ref{Figure3}). These variations stress the importance of taking into account the temperature dependence of the specific heat when measuring at cryogenic temperatures and make difficult the extraction of $\rho_{bd}$ at values of $T_{cryo}<70$ K. 
\begin{figure}[t]
\centering
\includegraphics[keepaspectratio,clip,width=1\columnwidth]{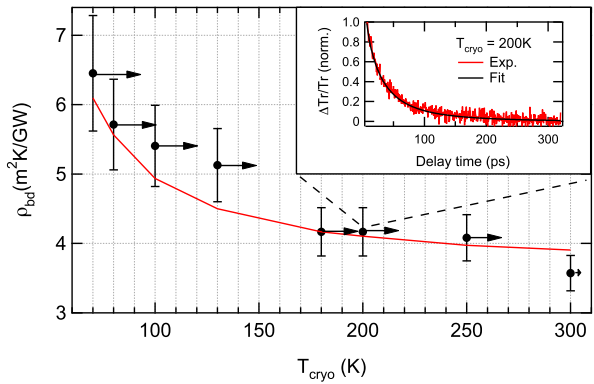}
\caption{(Color online) Kapitza resistivity $\rho_{bd}$ vs $T_{cryo}$ (black circles). The horizontal arrows indicate the temperatures spanned by the NP during the thermalization process. Plot of the function $AC_{p}^{-1}$, $A$ being a multiplication constant with dimensions ms$^{-1}$ (red curve). Inset: normalized transmission change (red curve) and its best fit (black curve) for the case $T_{cryo}$=200 K.}
\label{Figure3}
\end{figure}
\indent The Kapitza resistivity $\rho_{bd}(T_{cryo})$ increases by a factor of  two, spanning values from 3.2 to 6.5  m$^{2}$K/GW, as the cryostat's temperature decreases from 300 K to 77 K, see Fig.~\ref{Figure3}. The extracted value corresponds to a mean value of $\rho_{bd}$ over the NP's temperature excursion during the experiment (indicated by arrows in Fig.~\ref{Figure3}). Values for $\Lambda_{m}$ were found in the range 0.2-0.7 W/mK, comparable to the ones reported for thermal conductivity of glasses with similar compositions.\cite{Juve2009}\\
\indent  Starting from the general expression for the Kapitza resistivity \cite{Swartz1989}, and assuming both a frequency independent and/or frequency-averaged phonon transmission coefficient $\tilde{t}$, and group velocity $\widetilde{v_g}$, one finds $\rho_{bd}\sim(\tilde{t}\widetilde{v_{g}}C_{p})^{-1}$. This trend is experimentally retrieved in our data where $\rho_{bd}$ is found to roughly follow the temperature dependence of $C_{p}^{-1}$, see Fig.~\ref{Figure3}.\\
\indent In conclusion we measured via time-resolved all-optical nanocalorimetry the Kapitza resistivity of a 4.5 nm radius glass-embedded Ag nanoparticle in the temperature range from 300 K to 70 K. $\rho_{bd}$ increases monotonically from an ambient temperature value of 3.2 m$^{2}$K/GW to 6.5 m$^{2}$K/GW. The present findings constitute a benchmark for theories aiming at explaining the Kapitza resistivity in nanosystems, a fundamental issue for applications in nanomedicine and thermal management at the nanoscale.\\ 
\begin{acknowledgements}
We acknowledge Dr. Aureli\'en  Crut and Dr. Paolo Maioli for enlightening discussions and useful suggestions. This work was partially funded by grant D.2.2-2011 of Universit\`a Cattolica and the ¡ÈOpthermal¡É grant of the Agence Nationale de la Recherche. N.D.F. and F.B. acknowledge support from the Institut Universitaire de France and CNRS, respectively. Open Access publication was sponsored in the frame of Scientific Dissemination Grant D.3.1-2011 of Universit\`a Cattolica.
\end{acknowledgements}
\begin{widetext}  
Copyright (2012) American Institute of Physics. This article may be downloaded for personal use only. Any other use requires prior permission of the author and the American Institute of Physics.
The article appeared in Appl. Phys. Lett. \emph{100}, 011902 (2012); doi: 10.1063/1.3673559 and may be found at http://link.aip.org/link/?apl/100/011902.
\end{widetext}

\begin{thebibliography}{12}

\expandafter\ifx\csname
natexlab\endcsname\relax\def\natexlab#1{#1}\fi
\expandafter\ifx\csname bibnamefont\endcsname\relax
  \def\bibnamefont#1{#1}\fi
\expandafter\ifx\csname bibfnamefont\endcsname\relax
  \def\bibfnamefont#1{#1}\fi
\expandafter\ifx\csname citenamefont\endcsname\relax
  \def\citenamefont#1{#1}\fi
\expandafter\ifx\csname url\endcsname\relax
  \def\url#1{\texttt{#1}}\fi
\expandafter\ifx\csname
urlprefix\endcsname\relax\def\urlprefix{URL }\fi
\providecommand{\bibinfo}[2]{#2}
\providecommand{\eprint}[2][]{\url{#2}}

\bibitem{Hartland2011}
G. V. Hartland, Chem. Rev. \textbf{111}, 3858 (2011).

\bibitem{Juve2009}
V. Juv\'e, M. Scardamaglia, P. Maioli, A. Crut, S. Merabia, L. Joly, N. Del Fatti, and F. Vall\'ee, Phys. Rev. B \textbf{80}, 195406 (2009).

\bibitem{Plech2004}
A. Plech, S. Gr\'esillon, G. von Plessen, K. Scheidt, and G. Naylor, Chem. Phys. \textbf{299}, 183  (2004). 

\bibitem{Irsch2003}
L.R. Hirsch, R.J. Stafford, J.A. Bankson, S.R. Sershen, B. Rivera, R.E. Price, J.D. Hazle, N.J. Halas, and J.L. West, Proc. Natl. Acad. Sci. U. S. A. \textbf{100}, 13549  (2003).

\bibitem{El-Sayed2007}
P.K. Jain, I.H. El-Sayed, and M.A. El-Sayed, Nanotoday \textbf{2}, 18  (2007). 

\bibitem{Paasonen2007}
L. Paasonena, T. Laaksonenb, C. Johansb, M. Yliperttulac, K. Kontturib, and A. Urttic, J. of Controlled Release \textbf{122}, 86  (2007). 

\bibitem{Rini2005}
M. Rini, A. Cavalleri, R. W. Schoenlein, R. L\'opez, L. C. Feldman, R. F. Haglund Jr., L A. Boatner, and T. E. Haynes, Opt. Lett. \textbf{30}, 558  (2005). 

\bibitem{Swartz1989}
E. D Swartz and R. O. Pohl, Rev. Mod. Phys. \textbf{61}, 605  (1989). 

\bibitem{Stoner1993}
R. J. Stoner and H. J. Maris, Phys. Rev. B \textbf{48}, 16373 (1993). 

\bibitem{Cahill2003}
G. Cahill, W. K. Ford, K. E. Goodson, G. D. Mahan, A. Majumdar, H. J. Maris, R. Merlin, and S. R. Phillpot, J. Appl. Phys. \textbf{93}, 793 (2003). 

\bibitem{Siemens2010}
M. E. Siemens, Q. Li, R. Yang, K. A. Nelson, E. H. Anderson, M. M. Murnane, and H. C. Kapteyn, Nature Materials \textbf{9}, 26  (2010). 

\bibitem{Chen2001}
G. Chen, Phys. Rev. Lett. \textbf{86}, 2297 (2001). 

\bibitem{Palpant2008}
M. Rashidi-Huyeh, S. Volz, and B. Palpant, Phys. Rev. B \textbf{78}, 125408 (2008). 

\bibitem{Plech2004PRB}
A. Plech, V. Kotaidis, S. Gr\'esillon, C. Dahmen, and G. von Plessen, Phys. Rev. B \textbf{70}, 195423 (2004). 

\bibitem{Wilson2002}
O. M. Wilson, X. Hu, D. G. Cahill, and P. V. Braun, Phys. Rev. B \textbf{66}, 224301 (2002). 

\bibitem{Banfi2010}
F. Banfi, F. Pressacco, B. Revaz, C. Giannetti, D. Nardi, G. Ferrini, and F. Parmigiani, Phys. Rev. B \textbf{81}, 155426 (2010). 

\bibitem{Uchida1994}
K. Uchida, S. Kaneko, S. Omi, C. Hata, H. Tanji, Y. Asahara, A. J. Ikushima, T. Tokizaki, and A. Nakamura, J. Opt. Soc. Am. B \textbf{11}, 1236 (1994). 

\bibitem{Nelet2004}
A. Nelet, A. Crut, A. Arbouet, N. Del Fatti, F. Vall\'ee, H. Portales, L. Saviot, and E. Duval, Appl. Surf. Sci. \textbf{229}, 226 (2004). 

\bibitem{Giannetti2007}
C. Giannetti, B. Revaz, F. Banfi, M. Montagnese, G. Ferrini, F. Cilento, S. Maccalli, P. Vavassori, G. Oliviero, E. Bontempi, L. E. Depero, V. Metlushko, and F. Parmigiani, Phys. Rev. B \textbf{76}, 125413 (2007). 

\bibitem{Arbouet2003}
A. Arbouet, C. Voisin, D. Christofilos, P. Langot, N. Del Fatti, F. Vall\'ee, J. Lerm\'e, G. Celep, E. Cottancin, M. Gaudry, M. Pellarin, M. Broyer, M. Maillard, M. P. Pileni, and M. Treguer,  Phys. Rev. Lett. \textbf{90} , 177401 (2003).

\bibitem{Jaeger:book}
H. S. Carslaw and J. C. Jaeger, \textit{Conduction of Heat in Solids} (Oxford University Press, Oxford, 1959).

\bibitem{DeltaT0}
The value $\Delta T_{0}$ is evaluated solving $U_{V}=\int_{T_{cryo}}^{T_{cryo}+\Delta T_{0}} \! C_{p}(T)dT$, the electrons' contribution to the specific heat being negligible for the explored temperature range."

\bibitem{ThermalData}
$C_{m}(T)$ was taken from SciGlass database, $C_{p}(T)$ from F. Meads, W. R. Forsythe, and W. F. Giauque, J. Am. Chem. Soc. \textbf{63}, 1902 (1941).

\end{thebibliography}
\end{document}